# Stored and Inherited Relations

Witold Litwin[1]

**Abstract.** *The universally applied Codd's relational model has two constructs: a stored relation, with stored attributes only and a view, only with the inherited ones. In 1992, we have proposed third construct, mixing both types of attributes. Examples showed the idea attractive. No one followed however. We now revisit our proposal. We show that a relational database scheme using also our construct may be more faithful to reality. It may spare the logical navigation or complex value expressions to queries. It may also avoid auxiliary views, often necessary in practice at present. Better late than never, existing DBSs should easily accommodate our proposal, with almost no storage and processing overhead.*

## 1. Introduction

The universally applied Codd's (relational) model for a Database (Management) System (DBS), [C69] & [C70] has two basic constructs: a stored relation and a view. Both are named finite relations with atomic attributes only, in $1^{st}$ Normal Form (1NF) thus. A *Stored Relation*, (SR), called also a *base* one, or simply *relation* or a (relational) *table*, has *stored* (base) attributes (columns) only. A *view*, also called *Inherited Relation* (IR), has only the *inherited* attributes. These get values from SRs or from views through a stored statement of some data manipulation language (DML), usually a stored SQL query. In 1992, we proposed an additional construct. It was also a 1NF relation, but mixing the stored and the inherited attributes, [LKR92]. Examples showed the construct attractive. The idea seemed promising also for OODBs, *à la mode* in those times.

No one followed the lead however, to the best of our knowledge. Below, we revisit our proposal thoroughly, subsequently to [L6]. We call our construct *Stored and Inherited Relation*, (SIR). We qualify of *SIR-model* the data model incorporating SIRs. A view in SIR-model may thus also inherit from an SIR. We refer to Codd's model as to <u>S</u>tored <u>R</u>elation or <u>V</u>iew model, (*SRV-model*). We believe the reader familiar with the SRV-model and SQL in particular.

We show that SIR-model adds useful capabilities to all these of the SRV-model, preserved by definition. We propose some SQL extensions for SIRs. We restate the usual relational scheme design rules to include SIRs. It will appear, perhaps surprisingly, that a relational DB could often advantageously consist of SIRs only. We propose an implementation of the SIR-model over an existing DBS. It appears easy, with negligible storage and processing overhead. We hope our model entering the practice "better late than never".

Next section details the SIR-model. We discuss the basic concepts and the SQL extensions. We show that conceptual schemes with SIRs should be usually more faithful to the reality. The ER-model becomes rather useless. Queries may be free from the customary logical navigation through inter-relational joins. SIRs appear the first generally practical solution to this decades' old annoyance. Likewise, SIRs may spare complex aggregate expressions in queries, another old pain. Both troubles are unavoidable for an SRV-model database (DB) at present, unless the database administrator, (DBA), creates additional dedicated view schemes. Actually, at present, queries with aggregate expressions often require auxiliary views anyway. Managing multiple views is however cumbersome as well, enough to be practiced in general only when unavoidable.

Section 3 shows our restated relational schema design rules. We first generalize the NFs other than 1NF to SIRs. Next, we restate the Heath's and Fagin's theorems. The restated theorems create lossless decomposition into projections that are SIRs instead of the usual stored ones in SRV-model. The result avoids usually the logical navigation over the projections, while the present principles create one necessarily. More precisely, for the restated Heath's theorem, our decomposition is totally logical navigation free. For the restated Fagin's one, some need remains, likely occasional in practice.

---

[1] Université Paris-Dauphine, PSL Research University, CNRS, LAMSADE, 75016 Paris, France.  witold.litwin@dauphine.fr

Section 4 discusses the implementation of SIR-model over an existing DBS, as well as its storage and processing overhead. Section 5 draws the conclusions and overviews the future work.

## 2. SIR-Model

### 2.1 Overview

Figure 1 shows the SIR-model versus the SRV-model. As already discussed, the SR construct is the same for both models. An SRV-model view, i.e., an IR, is also a view for the SIR-model. The inverse is not true, as also stated. Next, as both constructs in SRV-model, an SIR is a finite subset of a Cartesian product of atomic attributes (columns) over some domains, subject to any algebraic or predicative operations, and aggregate or scalar functions, applying to 1NF relations. As we said however, unlike an SR or a view, some attributes of an SIR are the stored ones and some are inherited. A stored attribute (SA) has its values stored somewhere in the DB. The values of an inherited attribute, (IA), are calculated. As usual for an SR or a view, an SIR has a name and a scheme. The latter defines every SA and IA. We suppose every SA scheme to be a usual one, e.g., as in an SQL dialect. An IA in contrast has a specific scheme that we call *inheritance expression* (IE).

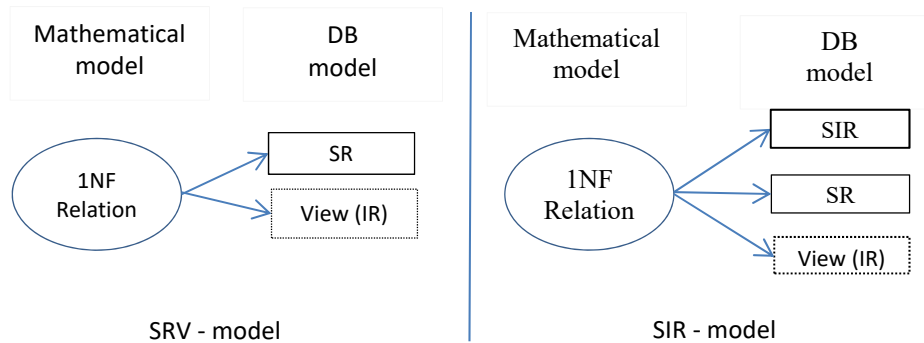

Figure 1 SRV-model versus SIR-model

An IE selects one or several IAs, e.g., in the Select clause for an IE defined using an SQL expression. These may be inherited from stored relations, views or SIRs. An SIR scheme may contain one or multiple IEs. Each IE defines a disjoin subset of IAs for the SIR. Every SA and IA in an SIR should have a unique proper name, i.e., without any prefix, like the usual relation name from which an IA to disambiguate could come from. Every IE selects some IAs and determines tuples of IA values using some relational selection expression, e.g., an SQL Select statement. As usual, we basically consider these values not materialized. Each tuple calculated through an IE is a sub-tuple of some tuple in the SIR with the IE, Figure 2. The SIR at the figure has multiple IEs. One sees there that each tuple of, e.g., $IE_1$, contributes to some SIR tuple with the sub-tuple with values of all the SAs and with the sub-tuples with values from all the others IEs. SAs are collectively named B at the figure for reasons we explain soon. A green rectangle represents each sub-tuple.

More precisely, for every SIR tuple *t*, we require every IE to calculate at most one tuple of IAs values to become a sub-tuple of *t*. For this purpose, every IE contains a specific join clause. This one matches attributes in relations the tuples the IE creates inherit from, i.e., named in SQL From clause of the IE, with some SAs or IAs in the SIR beyond the sub-tuple. Let R be an SIR, *I* an IE and R' = R / *I*. Since, every attribute of R' belongs also to R, we use in the clause R name to disambiguate an R' attribute name. We call the join *recursive*, since it defines R by referring to R. It can be any $\theta$-join, although an equijoin should be the usual one. For every SIR, the recursive join in at least one IE should match some SA. Also, for each R' tuple *t'*, at most one *I* tuple, say *t*, should match the recursive join. Tuple *t* becomes the sub-tuple of the R tuple with the matching *t'*. If for some *t'*, *I* does not find any matching *t*, we set the sub-tuple to the null *I* tuple, i.e., with the null instead of every IA selected by *I*. In other words, for every *t'*, the result of a recursive join clause is the join of *t'* with *t* or with the null *I* tuple. The result for any R' and *I* is thus a left outer join. For user's convenience, we allow nevertheless the clause itself to be an inner join or even to be implicit, as examples will show soon. Figure 2 shows the result also when null sub-tuples result. E.g., the bottom tuple contains the null $IE_1$ sub-tuple and the null $IE_2$ sub-tuple.

Next, let us write an SIR R as R (B, V) with B containing all the SAs, and V all the IAs. V consists thus of the columns labelled with IEs at Figure 2. For every R, we require B to be a stored relation for SRV-model, i.e., without duplicates. We call B *stored relation* of R or *base* of R. As every stored relation, each base has its own name that is R_B by default. Likewise, every SIR has one or more keys. We require all keys of B among these. We recall that for any relation, all attributes are functionally dependent on every key. For any key value of any R, we have thus for each SA and each IA, at most one (atomic) value, respectively stored or inherited. The rationale is, again, the requirement of 1NF for any SIR. Finally, we always have card (R) = card B. All this is also the rationale for the requirement we made above that for every R tuple, for every IE, its recursive join matches at most one IE tuple.

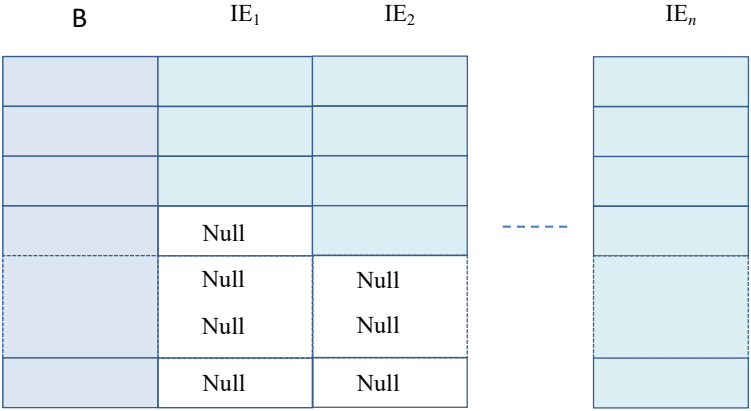

Figure 2. An SIR with multiple ISs. Stored values are grey, inherited ones are green. IEs actually bring respectively three, four and as many tuples as there are stored ones. V unions all the IAs.

We require also every IE to bear a name, unique within its SIR scheme. The IE name may be explicit or implicit as it will appear. It serves as the reference to its IE, as we'll show. If needed, we disambiguate IE names as usual for SQL.

We call *SIR DB* a DB conform to SIR-model. We suppose SIR DB schemes defined as usual through some DDL (Data Definition Language). Operationally, we consider from now some SQL-like DDL. We suppose the statements of this DDL extend to SIR-model some existing SRV-model conform SQL dialect. We call the latter the *kernel* for the resulting SIR-model dialect. We presume the kernel to be the dialect of some major DBS. E.g., Create Table for an SIR may extend the MySQL Create Table statement with the capability of defining an IE.

An IE may inherit from a stored relation, a view or an SIR. It may happen that an SIR or a view R1 inherits from an SIR or a view R2 and vice versa. By extension of the current terminology for views, we say that R1 and R2 are then in a *circular reference*. We forbid circular references in an SIR DB, like they are prohibited at present at any existing major DBS we are aware of. The DBSs detect them at the run-time, issue some SQL Error Code and abort the query. Our rationale for extending this policy to SIR DBs is that the basic processing scheme (BPS) of SIR DBs we detail in Section 4, supposes the implementation over an existing major DBS. That one in particular takes care of the actual query evaluation for the SIR DB. For any circular reference in the SIR DB, BPS would then create a circular reference in the DBS. The query to the affected SIRs or views in the SIR DB would consequently not get processed.

The straightforward way to avoid a circular reference in an SIR DB is to rewrite one of the IEs so it refers to the stored relations only. These can be bases of SIRs that the IE refers to. Also, they can be the origin of IAs in those SIRs. The latter applies to IAS views as well, including in present DBs. We illustrate the idea with examples soon.

From the IE description above, one could observe that the concept generalizes to possibly multiple attributes the one of a subquery in Select clause of a standard SQL query. The latter contributes to each tuple created by the query, with a single (atomic) attribute value, at most. A sub-query must also contain a kind of recursive join, between some attributes of relations the sub-query inherits from and some attributes other than the created one in the relation the sub-query contributes to create. This join determines for each value produced by the sub-query the query tuple this value enters. The IE similarly contributes with a single sub-tuple to each tuple of the relation being defined. But, that sub-tuple may be a composite one. Except for this difference and different purposes of course, the SQL expressions for both constructs may have the same syntax.  E.g. the expression (Select A From F Where F.J = R.J) could serve a sub-query of a query to some relation R or could serve an IE contributing to define some SIR R. In contrast, the expression (Select A1, A2 From F Where F.J = R.J) could serve an IE only. Both could generate null values, as if their inner joins were the half outer ones we spoke about.

We now illustrate all the discussed points with a motivating example. We come back to the general discussion of the DDL for SIRs afterwards.

## 2.2 Motivating Example

We reuse the biblical Codd's Supplier-Part relational DB, originally illustrating his proposal, as popularized by C.J. Date, [D4]. It is often named S-P DB in short. The example restates S-P using SIRs. We refer to the original scheme as to S-P1. We call the restated scheme S-P2.

Example 1. S-P1 models an enterprise with some Suppliers, Parts and Supplies. A supply contains some quantity of a part shipped by some supplier. Besides, some suppliers may be supplying anything currently, as well as parts may be not supplied at present. S-P1 conceptual schema consists of three stored relations named S, P, SP. This scheme is optimal for the so-called *relational design* rules, using NFs in SRV-model. In S-P1, SP respected the referential integrity by default. At present, one would need Foreign Key clauses. We suppose the default referential integrity for S-P2 as well, till we state otherwise. For every supply tuple in SP, the supplier tuple with the same S# must be in S and the part tuple with the same P# in P.

1. Figure 3 shows the S-P2 scheme. It is also optimal in SIR-model, as it will appear. Figure 4 shows example extensions of S-P2 relations, i.e., a possible content of each. The schemes and the contents of S and P relations are the original ones.  We use self-explaining statements to define every relation. We underline the primary key attributes, as usual. The S, P relations have SAs, but not IAs. They were stored relations in S-P1 and remain so in S-P2. SP keeps the original SAs with their values, i.e., these of S-P1.SP that are (S#, P#, Qty) to recall. These SAs form the base B of SP. B has its original primary key, (S#, P#). SP however has now also IAs. The left to right order of SP attributes follows the top-down and left-to-right in the IEs order of their definitions in the SP scheme. Also, B key is the primary one for SP as well. The original SP became now an SIR. All the IA of SP form V in our generic notation. The IA names and values are in Italics at the figure. The choice of IAs means that the Database Administrator (DBA) considers every property of a supplier and of a part, as also, *ipso facto*, of any supply. We examine the rationale for it later.

There are two IEs in SP-2 scheme, named I_S and I_P.  An SQL Select expression, supposed syntactically conform to some kernel SQL dialect, defines each IE. I_S inherits all and only attributes of S, except for S#. Its tuples inherit values (data) from the tuples of these attributes in S. Likewise, I_P inherits all and only attributes of P, except for P#. Both have renamed the source attributes CITY into proper names unique in SP, although only one renaming would obviously do.  The clause SP.S# = S.S# is the recursive (equi)join clause for I_S. As required for at least one IE in an SIR, I_S matches thus an SA in SP. SP.S# is here also the R' attribute we spoke about, i.e., outside all the IAs created by I_S. Same for SP.P#, respectively. For the first tuple of SP, with SAs S# = S1 and P# = P1, the recursive join clause in I_S matches I_S tuple (Smith, 20, London), since S.S# = S1 for this tuple. SP.S# = S1 does not match S# value in any other I_S tuple, as required. The obvious reason, for the similar result for every SP.S# value besides, is that S# is the key of S. The selected I_S tuple becomes the sub-tuple created by I_S in the first SP tuple.

```
S-P2 Scheme

Table S            Table P            Table SP

S#  Char,          P# Char,           S# Char,
SNAME Char,        PNAME Char,        P# Char,
STATUS Char,       COLOR  Char,       QTY Int,
CITY Char;         WEIGHT Char,       I_S (Select SNAME, STATUS, CITY As SCITY From S
                   CITY Char;           Where SP.S# = S#);
                                      I_P (Select PNAME, COLOR, WEIGHT, CITY As PCITY
                                        From P Where SP.P# = P#);
```

Figure 3  The S-P2 scheme.

```
S-P2 Content
Table S                                    Table P
S#      SNAME  STATUS  CITY                P#      PNAME   COLOR   WEIGHT  CITY
S1      Smith  20      London              P1      Nut     Red     12      London
S2      Jones  10      Paris               P2      Bolt    Green   17      Paris
S3      Blake  30      Paris               P3      Screw   Blue    17      Oslo
S4      Clark  20      London              P4      Screw   Red     14      London
S5      Adams  30      Athens              P5      Cam     Blue    12      Paris
                                           P6      Cog     Red     19      London
Table SP
S#   P#   QTY    SNAME  STATUS  SCITY    PNAME   COLOR   WEIGHT  PCITY
S1   P1   300    Smith   20     London   Nut     Red      12     London
S1   P2   200    Smith   20     London   Bolt    Green    17     Paris
S1   P3   400    Smith   20     London   Screw   Blue     17     Oslo
S1   P4   200    Smith   20     London   Screw   Red      14     London
S1   P5   100    Smith   20     London   Cam     Blue     12     Paris
S1   P6   100    Smith   20     London   Cog     Red      19     London
S2   P1   300    Jones   10     Paris    Nut     Red      12     London
S2   P2   400    Jones   10     Paris    Bolt    Green    17     Paris
S3   P2   200    Blake   30     Paris    Bolt    Green    17     Paris
S4   P2   200    Clark   20     London   Bolt    Green    17     Paris
S4   P4   300    Clark   20     London   Screw   Red      14     London
S4   P5   400    Clark   20     London   Cam     Blue     12     Paris
```

Figure 4  The S-P2 content. IAs are in Italics.

Likewise, I_P creates, for the first SP tuple, the sub-tuple (Nut, Red, 12, London). Together with the B sub-tuple (S1, P1, 300) and the I_S sub-tuple above, all form the (entire) SP tuple. Similar calculation determines all the other SP tuples at Figure 4.  Notice that IAs defined by I_S as well as these of I_P are functionally dependent on the key (S#, P#), as we required for IAs of any SIR. Notice also, that there cannot be an SP tuple with a null I_S or I_P sub-tuple, because of the referential integrity constraint.

Comparison of equivalent queries to S-P1 and S-P2 easily shows how the latter scheme spares the logical navigation unavoidable with the former, as we spoke about. Consider the client wish to see the names, IDs and quantities of parts supplied by Smith. Queries of this type are among the most frequent to countless DBs, molded upon S-P1 since decades. The two queries could be:

(Q1) Select P#, PNAME, QTY From SP Where SNAME ='Smith' ; /* Query to S-P2

(Q2) Select  SP.P#, PNAME, QTY From S, SP, P Where SNAME ='Smith' And S.S# = SP.S# And SP.P# = P.P# ; /* Equivalent query to S-P1

Any users should prefer (Q1) as less procedural. The joins between SP and S and P in (Q2) are examples of the *logical navigation* we spoke about, [MUV84]. As well-known, users usually distaste that one. Most find it cumbersome to write and many have often trouble to figure out its precise functioning. Especially, since, for the notational compatibility with the outer joins, the SQL standard advocates these clauses in even more procedural form in From clause, using explicit Join verb and parenthesis-based algebraic notation. For instance, here we would have:

S Joins (SP Joins P On SP.P# = P.P#) On S.S# = SP.S#.

Observe finally that (Q2) could be formulated differently, using a subquery for PNAME in Select clause, namely as follows:

(Q2.1) Select P#, (Select PNAME From P Where SP.P# = P.P#) As PNAME, QTY   From S, SP where SNAME ="Smith" And S.S# = SP.S# ;

The subquery is the same as for an IE defining PNAME alone for our SIR SP. Its clauses other than Select are the same as for I_P. Accordingly to what we stated generally before.

2. Suppose now that the referential integrity towards S and P is not mandatory for SP in to S-P2, so that some client may insert into SP at Figure 4, the tuple showing supplier S7 supplying part P10 in quantity 200, without yet S# = 'S7' in S and P# = 'P10' in P. None would produce any matching tuple. The resulting SP tuple would have thus two null sub-tuples. The proverbial simplest SQL query Select * From SP; for instance, would show for S-P2, all the tuples in Figure 4 and the tuple (S7, P10, 200, null…null).  Notice that for S-P1, the equivalent query would need the logical navigation through outer joins. That one has a well-earned reputation of being even more awkward than through the inner ones, [DD91]. Perhaps, that is why, e.g., in MsAccess it remains bugged since its earliest version.

3. The STATUS attribute in S-P1 is a stored integer. Imagine it calculated behind the scene, e.g., as the total quantity delivered by a supplier divided by hundred and rounded to its integer part. Thus, the supplier of 100 - 199 parts has status 1 etc. A supplier that does not supply any parts for the time being has null as status. If STATUS value changes, while being stored, the client must execute an Update statement. Also, the value would be inaccurate in the meantime. A modern client may find the issue moderately practical. If S-P1 is the scheme of an SIR DB, the DBA may alter S making it an SIR with STATUS being an IA through the following IE. No need for Update queries and the value read by any queries would be always up to date.

STATUS (Select Int ( SUM (QTY) / 100) From SP Where S.S# = S#);

Notice that, for every S#, STATUS produces at most a single value. The name of the IE is implicitly that of the IA it creates. Next, as required for a mono-attribute IE. Suppliers S1…S4 would have some STATUS, but not S5, not supplying anything currently. STATUS of S5 would be null, consequently. Next, observe that under SRV-model, S can only be a stored relation. If STATUS is to  be always up to date, one way is to dynamically calculate it as a sub-query in the Select clause of every query to STATUS, e.g., in the query:

Select S#, SNAME, CITY, (Select Int ( SUM (QTY) / 100) From SP Where S.S# = S#) As STATUS From S;.

Notice that for every S#, the sub-query produces at most a single value. The subquery is besides quite procedural. Writing it for a query selecting STATUS might be not liked by many. In addition, the solution does not work on popular DBSs for all usual queries. E.g., one cannot add to the above query neither the clause Order By STATUS nor Order By (Select Int…), i.e., the subquery defining STATUS, e.g., for the popular MsAccess[2]. S-P2 user may freely add the clause, in contrast. In practice, it is mandatory today to expand S-P1 (conceptual) scheme with a dedicated view scheme. The simplest one could be:

Create View Status As Select S#, Int(SUM(QTY)/100) AS STATUS FROM SP GROUP BY S# ;.

However, the immediate practical advantage of S as an SIR is that the simplest query Select * from S continues to deliver STATUS. With Status view, the logical navigation S left join SP On S.S# = SP.S# is necessary. Alternatively, Status view should be more procedural. It should inherit also all the other attributes of S and include the outer join clause. Each solution of an additional view scheme brings havoc with respect to the single SIR S scheme. SIR-model clearly alleviates here another fundamental limitation of the SRV-model. We come back to this issue in the next section.

4. Suppose now that the calculated STATUS should apply to S in SP-2. If DBA declared it using the IE above, STATUS would inherit from SP, but SP would also inherit from S through I_S. The IE STATUS would create the circular reference between S and SP. A way out is to observe that attributes STATUS inherits from SAs of SP only. Hence, they are all in the stored relation SP_B. The DBA may declare STATUS therefore as the IE:

STATUS (Select Int ( SUM (QTY) / 100) From SP_B Where S.S# = S#);

5. Having STATUS in S, we wish to add, as immediately following it, one more inherited attribute named RANK. For each supplier *s*, RANK (*s*) value is either one plus the number of Suppliers with STATUS higher than that of *s* or RANK (*s*) is set to null if STATUS (*s*) is null. The following IE does the job:

RANK AS (IIF(status is not null, (select count(*) +1 from S X where x.status > s.status), null)) FROM S;

Here IIF is the scalar function, e.g., of MsAccess SQL dialect. The IE is a value expression, perhaps surprisingly to some. Our syntax for such an IE follows DBA convenience reasons discussed in next section. Also, the recursive join is not an equijoin. Next, notice that the calculation of RANK requires that of STATUS. This is a constraint for query evaluation we revisit in Section 4. Finally, consider the simple query of obvious interest:

Select S#, SNAME, STATUS, RANK From S Order By RANK;

As for STATUS alone, to the best of our knowledge no current major DBS allows for a single equivalent query, even with complex subqueries and some logical navigation. The only solution is another auxiliary view defining RANK, inheriting from STATUS view. Altogether SIR-model would avoid here the burden of two auxiliary views and perhaps of a query still more navigational than the above one.

6. Consider finally that DBA of S-P1, finds that a conceptual property of every part, desired to be an additional attribute of P thus, following all the others in P scheme, should be the list of all the suppliers of the part, called SUPPLIERS and containing S#, SNAME. The list should be sorted first in descending order on quantity supplied, then the ascending one on SNAME. If a part is not supplied for the time being, SUPPLIERS should be null. IF P can be a stored relation only, i.e., under SRV-model, DBA wish is simply a dream. If S-P1 is an SIR DB, i.e., P can be an SIR, altering the original P with the following IE will make DBA happy. The LIST aggregate function in the IE casts for each supplier all the selected tuples into a single Char type value (a string thus), [L3].

---

[2] Perhaps surprisingly, MsAccess is the most popular relational DBS by number of licensees, allegedly in hundreds of millions.

SUPPLIERS (Select LIST (SP.S#, SNAME, QTY) From SP, S where P.P# = SP.P# And S.S# = SP.S# Order By Qty Desc, SNAME)

The result, e.g., for P1 would be P tuple:

(P1, Nut…London, (S2, Jones ; S1, Smith))

Without LIST aggregate, SUPPLIERS would not be an IE. The expression could inherit more than one tuple for some P tuples, e.g., for these with P1. Also, observe that if P was in S-P2, then P altered as above would create a circular reference. The way out is the reference to SP_B instead of SP, as above for STATUS. Observe that the resulting IE would be fine for the altered S-P1 as well. Observe, finally again, that the simplest SQL query Select * From P is now equivalent to that with two joins between S, P and SP at present, i.e., in S-P1 still under SRV-model. Actually, without SUPPLIERS, the equivalent query would be more procedural even with S-P2, saving the logical navigation however.@

## 2.3  DDL Statements for SIR-model

Create Table for SIRs defines SAs and IAs basically as above illustrated. We suppose the syntax for SAs is that of the kernel Create Table. Also, as mentioned for S-P1, we suppose the Foreign Key clauses possible for SIRs as well. Furthermore, Create Table for SIR-model may in particular define a stored (only) table, supposedly using simply the statement of its kernel dialect. More generally, every DDL statement for SRV-model enters the SIR-model DDL by definition in this way. As one wisely said, who does more should be able to do less, Figure 1.

For views, in particular, we suppose the reuse as is of the kernel's Create View DDL statement. The rationale is the absence of the recursive join(s) in a view. Reuse of Create View seems then easier to implement than some view scheme only specific syntax in Create Table for SIR-model. We come back to this issue in Section 4.

The other usual SQL DDL statements are, we recall, Alter Table, Drop Table, Drop View and Create Index. With respect to the generalization to SIRs of Alter Table, we suppose that it keeps its kernel dialect clauses Add, Alter or Drop with all their capabilities, to operate on any SAs. We add the extension specific to SIRs, namely that Alter applies also to the IEs. The alteration may add an IE, after or before any specified SA or IA or after all the attributes, by default. It may also replace or drop an existing IE. This may transform an SIR into a stored relation or vice versa. However, we prohibit Alter to drop any attribute serving a recursive join. In particular, Alter cannot thus transform an SIR in a view.

Then, the extended Drop Table simply drops as usual the definition and eventually the content of an SIR. The operation should not of course violate the referential integrity. It might thus be required to cascade to other SIRs or get aborted if a violation results. Furthermore, as for any kernel we are aware of, if a view scheme should get altered, one should use Drop View followed by Create View statement. But, for SIR-model specifically, a view should also have potential to evolve into an SIR, i.e., get some SAs. We suppose then the Drop View followed by Create Table. Finally, we suppose that Create Index statement for an SIR applies naturally to its base only, reusing the syntax of the kernel one.

Example 2.

1. Suppose that in S-P2 WEIGHT is expressed implicitly in pounds, as apparently in S-P1 originally. Alter P by appending IA WEIGHT_KG converting it to kilograms and IA WEIGHT_T converting it further to tons. For application dependent reasons, WEIGHT_T should precede in the scheme WEIGHT_KG.

Alter Table P Add After WEIGHT  WEIGHT_T As ( WEIGHT_KG / 1000),
WEIGHT_KG  As (Round (WEIGHT / 2.1,1)) ;

These IEs would logically follow WEIGHT attribute in P, hence precede CITY. They inherit from P only. We qualify them thus of *self-inheriting*. We dispense a self-inheriting IE defined by a value expression from the full IE Select expression syntax. Nevertheless, one could define the self-inheritance alternatively in that way as well, e.g. as the IE:

I_W (Select  WEIGHT_KG / 1000 As WEIGHT_T, Round (WEIGHT / 2.1,1)  As WEIGTH_KG From P X, P Where X.[P#] = P.[P#]);

Main SQL dialects in fact already allow for both IAs, as so called *virtual*, *computed, generated*… columns. We come back to this practice in next section. The rationale for our basic syntax for an IE being a value expression is its compatibility with main dialects, for the DBA convenience. E.g., for SQL Server or MySQL, to append to P our WEIGHT_KG as a virtual attribute, the DBA would apply the same syntax as we would do for self-inherited IE WEIGHT_KG, namely: Alter P Add WEIGHT_KG  As (Round (WEIGHT / 2.1,1));.

2. We alter S in S-P2 by replacing STATUS with the inherited one we discussed.

Alter Table S Alter STATUS As STATUS (Select Int (SUM(QTY)/100) FROM SP_B WHERE S.S# = S#);

Notice that with P and S altered as above, S-P2 has only SIRs.   Below, we refer to the DB with P and I_P altered with WEIGHT_T, WEIGHT_KG and with STATUS IE as to S-P3.

3. We wish SP to implicitly inherit also eventual alterations of P, adding or dropping attributes there. The IE I_P does not do it. We suppose therefore that the usual SQL (Klein's) operator '*' in the dialect used, supports also the syntax */A or */(A1,…,An). The Select list that '*' generates, does not contain the attributes after '/'. We alter the SP scheme in Figure 3 as follows, replacing the current I_P with a new one.

Alter Table SP Alter I_P As I_P_ALL (Select */P.P# From P Where SP.P# = P.P#);

Notice that this Alter refers to I_P by its name (only). It replace I_P with a new name besides. Now, if one alters P as in (1) above, SP inherits its new IAs automatically. As usual for '*' operator, the '*/' may be thus more convenient than an explicit Select list, especially if the latter is long. However, the default inheritance is not for all applications, as widely known.

4. The DBA of S-P3 decided that a conceptual property of every supply, of obvious interest to every client, is the total weight of the parts delivered, expressed according to each of the three units of measure above. Also, the DBA decided that these values should precede any IA already in SP.  The following Alter Table statement may then respond to the DBA wishes:

Alter Table SP Add Before SNAME  I_SPWEIGHT (SELECT WEIGHT*qty As SPWEIGHT, SPWEIGHT_KG*qty As SPWEIGHT_KG, SPWEIGHT_T*qty As SPWEIGHT_T From SP X where s# = SP.s# and p# = SP.p#) FROM SP;

Notice that this IE refers to IAs already in SP. The rationale is purely didactic. It would be about as simple to use the originals in P. In SRV-model, one choice for the DBA would be at least one more SA in SP, with its storage cost, some way to calculate each stored value behind the scene and necessity updating it if the supplied quantity changes.  The only other choice would be an additional dedicated view scheme. Both results would be clearly more cumbersome than our altered SP SIR-scheme.@

## 2.4   Data Manipulation

Any SIR is a 1NF relation, by definition. The relational algebra operators of SRV-model operate on 1NF relations as defined by their mathematical model, Figure 1. Whether an attribute is a SA or an IA is immaterial to the operators. Each applies thus as is to SIRs as well. One may project, select or join thus any SIRs. The same holds for any SQL Select statements. Including these with value expressions, scalar and aggregate functions, the special clauses: Top k, Group By, Order By…

In short, SIRs do not require extensions to any current DML statements. It may nevertheless be practical to extend the '*' semantics as discussed in (3) above. For a modification of an SIR, i.e., the SQL Insert, Update or Delete statement, each continues to act for an SA and for an IA as it would act on stored relation or a view with such an attribute. The statement concerning an IA may thus succeed or fail, depending on whether a view update propagates. More precisely, an Insert statement creates as usual any SA value. As in our motivating example, we may require the Insert to commit only if it was also able to successfully calculate all the IAs.  The statement may also insert the value of an IA, provided the update may propagate to the source SAs. Create Table may or may not allow it for the referential integrity paths. If the modification cannot propagate, the entire statement should fail, e.g., if an IA is inherited through a value expression or a scalar or aggregate function. Same occurs for Update and Delete statements. The latter deletes as usual physically from the DB all the selected values of the SAs. It also deletes from the SIR the values of all the selected IAs, although conceptually only, of course.

Example 3.The simplest for SP SQL Select statement Select * From SP would show all the SP values, of all SAs and of all IAs in Figure 4. Supposing MS Access SQL as the kernel dialect, would make the statement Insert SP (select 'S4' as S#, P4 as P#, 100 as QTY); adding the tuple with these (stored) values and with all its IA values of I_S and of I_P virtually.  The statement Update SP set QTY = 250 where S# = 'S1' and P# = 'P1'; should normally succeed, updating one stored value. However, the statement: Update SP set QTY = 250, CITY = 'Paris' where S# = 'S1' and P# = 'P1'; may succeed iff a change to CITY value propagates to S.  To authorize it requires some thinking. The side-effect would be the city change for every other supply by S1, what may surprise. Next, an update of STATUS of SP may succeed, provided it propagates upward to S. But if it is the IA STATUS above defined, any update to must fail.  Finally, the statement Delete SP Where S# = 'S1'; would erase as usual physically from the DB all the values of the stored attributes in the selected tuples in Figure 4. Conceptually, it would also delete all their inherited values.

## 2.5   Utility of SIRs

SIRs enhance the usability of the relational model. For every stored relation B in a conceptual scheme of any DB under SRV-model, an SIR, say R, may have B as base and additional attributes. As these are IAs, they do not change any properties of B. Especially, B keeps its storage requirements, its normal form and thus the related absence of anomalies, as we will discuss. The obvious theoretical advantage of R is then the potential for a richer conceptual scheme of the "real" object that B and R are intended to model, e.g., a supply in S-P1 versus S-P2. The practical advantage may be that queries involving B and requiring the logical navigation, despised by many clients, may have equivalents addressing R without this pain. Similar gains may concern queries necessarily complex, i.e., highly procedural, when addressing B. Finally, queries to R, may avoid cumbersome auxiliary views required for equivalents having to involve B. It may be impossible to formulate the equivalent addressing B without such a view or the view may at least spare the logical navigation or complex expressions to the equivalent for B of the query to R, outright without these complications.

The discussion of S-P1, S-P2 and of S-P3 already illustrated these claims. As we said already, S-P1 is a template for countless DBs. The benefit from SIRs generalizes accordingly.  To recall the examples in the nutshell, for S-P1, with its three relation schemes, a query including any of IAs in SP of S-P2 or S-P3, would require the logical navigation. Likewise, only queries could provide the calculated values of STATUS, RANK, SUPPLIERS, SPWEIGHT… These queries would be rather procedural, being disliked by many as well. Cherry on the cake, some queries, e.g., with ORDER BY, would require also auxiliary views, as we discussed. Finally, under strict SRV-model, i.e., without virtual attributes, like, e.g., under MsAccess, only queries may calculate WEIGHT_KG and WEIGHT_T. The repetition of same value expressions in different queries is also likely to annoy many.

For S-P1, the only way out of all the above troubles at present is the creation of additional view schemes where at least the IAs of S, P and of SP in S-P2 or S-P3 become IAs as well. A practical solution is likely to contain three views at least.  One, say S', should define IAs having all the values of all the attributes of S in S-P3. Then, a view P' should do similarly for P and, finally, a view SP' for SP. For the same queries as to S-P3, one has now six relation schemes at least to deal with. This is clearly more cumbersome than managing the three schemes in S-P3 only.   E.g., at least since every scheme needs its proper unique name.

Actually, the logical navigation or complex queries versus auxiliary views is decades' old dilemma. The dilemma amounts to the proverbial choice between Scylla and Charybdis, [M4].  SIR-model appears the first generally practical way out. In particular, since SIRs embed two already popular view-saver practices.  The one of virtual attributes, amounting to simple arithmetic value expressions in self-inheriting SIRs, we hinted to. The other, sometimes qualified of *implicit joins* automatically adds joins to queries. However, popular DBSs, e.g. SQL Server or MsAccess, limit the practice to QBE interface, although they generate SQL internally, as we discuss soon. SIRs corrects finally conceptual modelling pitfalls of the SRV-model, well-known since Codd's proposal. These led to the popular Entity-Relationship model (ER), [C76]. ER is clearly useful for SRV-model for many, but basically useless for DBs with SIRs.

As we finally mentioned, the SRV-model is a strict sub-model of the SIR-model. The trivial condition to stay with the current model, under a hypothetical SIR-model enabled DBS, is simply to refrain from SIRs. Switching to the SIR-model is safe thus. No loss of any current capabilities of a relational DB may result from.

Example 4. Figure 3 and Figure 4 illustrate that if a DBS uses the SIR-model and if for any reasons, we wish S-P1 DB only, it suffices to drop I_S and I_P from S-P2 scheme. Any queries to S-P1 under SIR-model, amount then to these under SRV-model only.@

We'll now justify our claims more in depth. We start with the conceptual modelling through SIR-model versus that using SRV-model (only). We continue with the avoidance of the logical navigation, of complex queries or of the auxiliary views. Finally, we address our claim about the virtual attributes and implicit joins.

## Conceptual modelling

For the conceptual modelling, the known goal for a relational DB is to possibly (1) model in the DB conceptual scheme the properties of all the "real" objects wished by most users, (2) have all these properties in the stored relations only, in the smallest possible number. That is why we use n-ary relations and not, e.g., the once popular binary ones [A74]. A stored relation is not intended however for the calculated values. These are supposed in relational views or dynamically calculated in queries. A view in turn does not have any stored attributes. SRV-model cannot fulfill its conceptual modelling goal for the objects with properties of both types.

The SIR-model shares the goal, except for the restriction to stored relation in (2). SIRs by definition can model therefore objects with properties of both types.

Example 5. In S-P1, if STATUS is no more a stored attribute, it cannot be in the stored relation S. A query or some view, say Status, must provide its values instead. A stored relation alone does not suffice for the conceptual scheme of object named S. No such pitfall obviously for S scheme in S-P3@

Another and by far more famous pitfall concerns SP relation in S-P1. Recall that the relational model illustrated with S-P1 was an instant hit. Most folks had no problems with S and P schemes. One could easily imagine practical use of these data, despite the inherent simplicity of 1NF and even the limitation we just pointed out. In contrast, almost immediately, many did not agree with SP scheme, despite its stored attributes only. It's hard to imagine an actual supply with properties modeled by the three SP attributes only. At least the part and supplier names are a practical must.

Some could think that the obvious way-out was simply to store in SP also all the other attributes of S and P. Codd has pointed out however already in his original article that it would not be a wise idea. The so-called *anomalies*, initially called *strong redundancies*, would follow, because of violations of the NFs. We recall that, for one, an important storage overhead could appear. Next, the same value could need many re-inserts, with evident risk of error at least.

A user could be nevertheless happier with additional attributes despite the anomalies. However, Codd conjectured that most users would be rather unhappy. According to earlier work, e.g., on the Codasyl model and ANSI-SPARC reference architecture more generally, the conceptual scheme should be by definition the one the most acceptable for the commonwealth of the DB users. Codd postulated that the relational conceptual scheme should be therefore the one formally minimizing the storage for the set of stored relations free of anomalies, [C69]. Usually, a single relation cannot fulfill this goal. SP for S-P1, but also S and P, seemed, after all, the choice the most conform to these criteria, as widely known. The conceptual modelling insufficiencies for some users should be compensated somehow again by additional dedicated view(s), of S, P and of SP in S-P1 in particular.

But this postulate constitutes the other of the discussed limitations of the SRV-model. The model clearly annoys likely numerous commonwealths of DB users needing conceptual attributes that cannot be in stored relations without anomalies or being inherited only, as we just pointed out. SP in S-P2 is free of this trouble, while the "troubled" SP in S-P1 is the best choice under SRV-model. This is the case of S and P as well, when extended to SIRs in S-P3, by the inherited, but conceptual attributes we dealt with in the examples, the last one for S in particular.

In sum, for unknown reasons, Codd postulated that each set of stored attributes to minimize the DB storage for, should be a stored relation, called, besides, also a *base* relation, given its goal of the conceptual modelling. The SIR-model shares the Codd's goal of the minimal storage. But, to form a relation, the resulting stored attributes, may be accompanied by those not-impacting the storage, i.e., the IAs. SIRs become the "base" relations, instead of the Codd's stored ones only. In other words, Codd's postulate continues to apply, but is restricted to the SIR bases only.

It's worth recalling also that a heated debate followed the S-P1 scheme proposal almost immediately, pointing out to the pitfalls of SP especially. The popular ER model we already hinted to, resulted from. The ER diagram was recommended as the actual conceptual scheme of a relational DB. The actual (stored or base) relation schemes should model then that one, as a kind of internal logical scheme. For S-P1, S & P relations modelled ER-entities. SP modelled in contrast "only" the ER-relationship between. The three attribute could suffice then. However, convincing folks that a heavy box of parts they face is "only" a relationship seems not obvious. Likewise, the question whether a marriage is an entity or a relationship never got a convincing answer. SIRs avoid such esoteric troubles. The need for the ER-model largely if not totally disappears. Especially, since the trouble with the calculated only properties remained anyhow.

Finally, it is the common knowledge that the concept of inheritance with its sub-classes, sub-types, sub-tables… is "the heart" of the object oriented (OO) programming and systems. It got also incorporated in 90ties into the, so-called, object-relational database systems. The popular open-source PostgreSQL DBS is the most prominent survivor of this trend, [S96], [P]. The inheritance in PostgreSQL has a dedicated DDL clause, called INHERITS, in its Create Table for a sub-table. The clause incorporates into the sub-table the entire scheme of the super-table. The inherited scheme is expanded with the schemes of the attributes specific to the sub-table. The tuples in the sub-table, e.g., in (state) Capitals table in the flagship inheritance example in PostgreSQL tutorial, are not supposed to be also in the super-table, i.e., in Cities table in this example. Each tuple of the latter models only a city that is not a state capital. The SQL DML semantics was consequently modified to realize a default UNION of the content of the super and of its sub-table(s), when a Select query addresses the former, e.g., for the selection of some attributes of any city using the SQL clause From Cities. To avoid such a union, the explicit keyword ONLY is necessary in the Select expression. Notice finally that one has to know also upfront whether a city to insert is a capital or not.

This implementation of the inheritance concept, including the modification of the SQL DML semantics, is unique to PostgreSQL to the best of our knowledge. It clearly differs from that for an SIR. First it concerns only schemes and only of all the stored attributes that becomes the schemes of stored attributes in the sub-table. For an SIR, an IE concerns both schemes and tuples. As for views, the source can be SAs, but also IAs, while the result is only IAs. The SQL DML semantics remains also unchanged. For the PostgreSQL example, since no tuple of Capitals is supposed to be in Cities, while Capitals is still supposed to have the same stored attribute schemes as Cities, SIRs would be of no use.

The rationale for it is that SIR-model has a different conceptual model of inheritance. For PostgreSQL example, the Capitals table could have the same intention, but Cities should be a stored relation for all the cities. To avoid any relational anomalies in Capitals, the only shared stored attributes should be, as usual, the primary key of Cities, becoming the foreign key in Capitals. This is actually the basic interpretation of inheritance for the declaration of sub-tables in MsAccess, we discuss soon. It is also advised for modeling inheritance under SQL Server (see any SQL Server Tutorial). In our case, an IE in Capitals scheme that would become then that of an SIR, could inherit consequently from the other attributes of Cities, but as IAs only. It could inherit from all of them, mimicking PostgreSQL example scheme of Capitals, or only from some. The latter capability could be useful if Cities had for instance also an attribute indicating the distance of a city to its state capital. Finally, an IE could in particular transform some inherited values. Examples are easy to imagine. The PostgreSQL INHERIT clause does not provide any equivalent capabilities.

**Logical Navigation, Complex Queries, Auxiliary Views**

Our next claim has two reasons, occurring in probably any relational DB at present. First, the SRV-model relational design principles that we recalled and also address more in depth in next section, often lead to queries to several relations at once. The mandatory logical navigation through natural inter-relational joins results from. It annoys many, [MUV84]. Likewise, as we pointed out in short already, many queries need results of complex value expressions. Possibly with aggregate functions, GROUP BY, subqueries in Select list or in the Where clause etc. Many users are simply unable to formulate such queries. Additional view schemes, shielding the logical navigation and complex calculus are the only fix at present. More schemes is however also a hassle, as we have shown. SIRs avoid both facets of the dilemma.

Example 6. For our calculated STATUS, the basic SRV-model solution is to create S without it and calculate STATUS in every query needing it. Some S-P1 clients could nevertheless find the value expression for the STATUS too complex for their taste. Under SRV-model, the only fix is an additional view with some or even all the attributes of S and the STATUS calculation, hiding its complexity, as we discussed in Example 1.3. The drawback is now however two relation schemes to manage. Where, in addition, in the latter case, the S scheme ends up useless for queries. The SIR S in SP-2 with the inherited STATUS avoids all these troubles.

The queries (Q1) and (Q2) from Example 1 illustrated how SIR-model may avoid the logical navigation. The only fix hiding the latter under SRV-model would be an additional view with all the attributes involved in the queries, e.g., say V2 scheme:

Create View V2 As Select SNAME, P.P#, PNAME, QTY From S, SP, P Where S.S# = SP.S# And SP.P# = P.P# ;@

More generally, it is easy to see that for any SIR R = (B, V), there is under an SRV-model a couple (B, V'), where B is a stored relation and V' is the view of all the attributes of R. V' inherits thus all the attributes and values of B as they are. It also inherits all the attributes of V, through an expression combining all the IEs of R. It is quite easy to see that such an expression always exists. It may be however rather complex for several IEs. A hierarchy of views may be a practical necessity. We actually apply this approach for an implementation of SIR-model over an existing major DBS, e.g., MySQL, proposed in Section 4. To avoid the logical navigation and complex queries, SRV-model may thus in practice at least double the number of relational schemes sufficient for same queries under SIR-model. In the same time, SIR-model schemes used as client interface do not induce any sensitive performance overhead, as it will appear. All these theoretical and practical properties of SIR-model constitute thus a clear advantage for the relational conceptual modelling over the current practice.

**Virtual Attributes**

The next claim was that SIRs generalize in fact some popular practices already beyond SRV-model, as originally defined. The first one is a view-saver usually called *computed, dynamic* or *virtual* attributes or columns. The concept appeared in 80ties. Major DBSs, Sybase first, picked it up rapidly and still use it. They do it without however, regretfully perhaps, of some research results, [LV86]. Virtual attributes are not stored, but only inherited through simple arithmetic calculations over stored or other virtual attributes in the same relation. A stored relation altered with virtual attributes becomes a self-inheriting SIR.

The IAs WEIGTH_KG and WEIGHT_T in Example 2 are virtual attributes. Self-inheriting SIRs, limited to basic arithmetic value expressions, are thus in fact already widely applied. Virtual attributes are an add-on to SRV-model, as they create SIRs. Strict observance of that model would require the dedicating view with such attributes. Such views would be always computationally sufficient. The concept is thus only a view-saver, intended to enhance the usability. The conjecture appears true. As said, major DBSs propose the concept for decades. Our examples have shown that SIRs in general provide for by far more extensive calculus capabilities, including the inheritance from multiple relations. These appeared helpful at least for S-P1 becoming our S-P3. S-P1 being the template today for countless actual DBs, we expect similar magnification of these capabilities of SIRs.

**Implicit Joins**

Research proposed different ways to avoid the logical navigation. The *universal* relation idea we recall in next section was the basis for one group of proposals. The *implicit joins*, sometimes called now also *automatic*, were an alternate [L85], [LSW91]. The universal relation, despite strong excitement, [M04], did not make to popular DBSs. The implicit joins did, e.g., into SQL Server & MsAccess. Again, the industrial versions limited the research results. The 2016 MsAccess version seems the most extensive up to now. Strange enough, the two MS systems use implicit joins only for the QBE interface. The graphical queries with implicit joins translate to SQL, with the joins added. In a QBE query graph, the implicit joins are directed or undirected arcs. They pop up once one selects the query relations, represented as the graph nodes. Alternatively, through the definition of so-called *sub-tables*, the implicit joins help 4GL forms, called *data sheets*. We detail these terms soon.

The query arcs are derived from directed or undirected arcs, called ambiguously *relations* between tables in a specific diagram of the DB scheme and of views, termed Relationships. The arcs are optionally manually dragged between the diagram nodes that are boxes representing the actual relations, called tables. These may be stored tables or views. One may declare the referential integrity when appropriate and the type of join to be implicit in queries. This can be an inner equijoin (default) or a half outer-join, translated to left or right in SQL. MsAccess may also automatically propose for the query arcs that are not appearing in the Relationships diagram, provided the DBA permission. In fact this was the initial purpose of implicit joins. In practice, the join attributes, must share the name then and one must be a primary key. The MsAccess automatic join is always an inner equijoin. The attributes involved may be composite. The SQL query generated from can be strange then however, especially for outer joins. The reasons are perhaps clear for Microsoft.

If an arc primary-foreign key exists between two tables, then the table with the foreign one may also automatically become a *sub-table*, we just spoke about. Sub-tables generated in this way seem the MsAccess application of the concept of inheritance, equivalent to that of References clause in SQL Server. One can also declare a sub-table manually within the so-called *properties* of the super-table. The sub-table is chosen by name and by declaration of an arbitrary atomic attributes per table as implicit join attributes, to select sub-tuple(s) of each super-tuple. Assuming the super-table at the left, the semantic is the implicit join is that of the left equijoin. In this way, e.g., one may declare S a sub-table of SP, to "inherit", which means make available for display by a mouse click in practice, all the details of a supplier for any S# value chosen in SP. MsAccess then automatically chooses SP.S# and S.S# for implicit left join. For unknown reasons, a table may have only one sub-table. If there are several arcs, as it would be for SP, and no manual declaration, one of the arcs is mysteriously preferred. Creation of sub-tables does not avoid the logical navigation in ad-hoc queries. It only let the sub-table tuples to be displayed, as we mentioned, either in as a sub-form of the 4GL form of the super-table, or in the specific view of the super-table, called data sheet view, we mentioned as well. In the data sheet of SP, for instance, there would be one line for every supply. Right under each such line, one could also display through the implicit join, an on-demand line with all the data of the supplier in S.

In this way, the declarations of sub-tables and the arcs of the relationship diagram, spares at least partly the logical navigation. The diagram avoids preexisting views of all these tables, perhaps even the universal one that would be the only way toward the goal under SRV-model. The implicit joins in general are intended as view-savers, e.g., of join clauses between S and SP or SP and P. The implicit joins generated by the MsAccess arcs share the intention. The SIR-model aims at similar capabilities as we showed, but, as for virtual attributes, exceeds potentially the current limitations. E.g. through its IEs, SP can be trivially dealt with as having two sub-tables, S and P, what is impossible for MsAccess at present. Likewise these IEs have the view-saving capabilities for complex values expressions that we discussed. The implicit joins were not even intended for. Summing up, the SIR-model usefully generalizes also that popular practice. Revealing a single umbrella for both discussed practices, what we claimed as well.

The common umbrella brings finally one more practical advantage worth mentioning. Our examples showed that if there is a choice for an SIR, say R again, multiple IEs should be usually preferable. To avoid the discussed troubles without SIRs, i.e., under the SRV-model, one may create for each IE in R a somehow equivalent partial view. One may stay with these views or, often better, combine all these for each R into one, say *full*, equivalent to R for any queries. These are in fact basically the various views practiced today we spoke about abundantly. As already stressed too, the approach is however rare. The manual practice of generating views is often procedural and error prone. SIR-model implemented over an existing DBS eliminates this trouble. The algorithm we define for this purpose in Section 4 generates all such views automatically.

## 3. SIR-model Schema Design

The relational scheme design rules have been studied for the SRV-model only. The overall goal was to avoid the anomalies. We now extend these rules to SIRs, for the same goal. We continue with S-P2 as the motivating example. We first restate the NFs. Next, we restate the Heath's and Fagin's theorems. The restated theorems generate the same lossless decompositions, but with SIRs instead of the original relations. The benefit is the total avoidance of the logical navigation, necessarily generated by the original decompositions, for the restated Heath's theorem and partial for the Fagin's decomposition.

## 3.1 Normal Forms

The basic design rule for a relational DB scheme under SRV-model is the respect of the normal forms (NFs). We recall that these are 1-3NF, BCNF, 4-5NF. Any relation in 5NF is in 4NF that is in BCNF etc. Every relation in SRV-model is by default in 1NF we also recall. Next, relations in 4NF that would not be in 5NF are rare, what makes BCNF and 4NF the most useful in practice. E.g., SP (S#, P#, QTY) in S-P1 is in BCNF, while SP' (S#, SNAME, P#, QTY) with stored attribute SNAME would not be. We'll give examples of 4NF later. Each NF eliminates some of anomalies we already signaled. E.g., SP' would need to store SNAME redundantly. Also, SNAME update could erroneously create two different names for same supplier. This could contradict S, where SNAME is anyhow already. Using SP instead, avoids the trouble.

First, recall now that any SIR is in 1NF by definition. Hence no need to restate this NF. The other forms have to be restated for SIRs. Observe in this context that the above anomalies of SP' would not exist for a view SP'. We therefore state that an SIR R (B, V) is in *i*NF or BCNF, iff B is in *i*NF or BCNF.

Actually, since R can have null values that were not in the original Codd's model, we implicitly consider as usual today that NFs apply to relations possibly with as well, e.g., as formally in [JS90].

Example 7. SP in S-P2 is in (extended) BCNF and 4NF, as well as in 5NF even. Indeed, the projection SP [S#, P#, QTY] on all and only stored attributes conforms to these NFs. Same happens, trivially, for S and P in S-P2. However, as mentioned, the stored relation SP' (S#, SNAME, P#, QTY) would not be in BCNF. But, an SIR SP' with IA SNAME in turn, would be. More generally thus, if, for any reasons, SNAME or any other IA in SP in S-P2 was rather a stored attribute, SP would cease to be in BCNF etc.@

## 3.2 Schema Design

We recall that at present, i.e. for a SRV-model DB, this process aims on a relational DB the (conceptual) scheme with possibly least number of relations free of anomalies. Usually, it means that every relation has to be proven as in 4NF or as at least in BCNF. The former need occurs if a relation present a (non-trivial) multivalued dependency (MVD). The latter, by far more frequent, characterizes schemes with the functional dependencies (FDs) only. The least number of relations means the grouping of all attributes functionally dependent on the same one(s) into possibly one relation, with the latter as the primary key. Possibly means here the respect of a myriad of other less or more fuzzy criteria, e.g., not "too many" null values for some attributes.

Designing a scheme is furthermore usually a many-steps process. Ideally, we start with the attempt of a single *universal* stored relation, say U, for the entire DB. U avoids the logical navigation entirely, as all the attributes are in. Unfortunately, chances for U in 4NF are zilch in practice. We usually perform then a decomposition of U into projections, i.e. we suppose that the DB consists of these projections instead. The decomposition must be *lossless,* producing the projections whose equijoin equals the decomposed relation. Any projection may end up proven in 4NF or proven in BCNF and free of any MVDs. It is then in 4NF thus as well. Or, a projection may not end up so. We decompose any such projections again. We continue, until all projections are anomaly-free.

As known, the two most used decomposition theorems are Heath's and Fagin's ones. The former may help with annoying FDs. The latter removes MVDs. Each theorem decomposes a relation into two projections. The resulting scheme has the least possible number of normalized relations for the DB, i.e., is of the smallest size and the optimal one in this sense. Actually, as only a few seemingly know, in presence of both MDs and FDs, Fagin's theorem must serve first. Otherwise a *sub-optimal* decomposition may result, meaning the scheme with more stored values than otherwise needed in a scheme nevertheless optimal in the sense we just defined. Even otherwise, there may be several decompositions that are all optimal in the discussed sense. So-called *independent* projections are preferable. Their known advantage is the preservation of the FD-cover. Rissanen's theorem testing the independence of the chosen projections may help.

We now generalize these principles to the SIR-model, i.e., U and the projections may be SIRs. Such schemes were out of scope of the original methodology, of course. In other words, even U may contain IAs, e.g. the aggregate ones we showed. For FDs and MVDs used for the decompositions, we nevertheless originally assimilate all these IAs to SAs. We apply to the projections the restated NFs. Then, in contrast, we consider any IA again as is. For the Heath's and Fagin's theorems rested for SIRs, our goal is that the decomposition of an SIR, say R again, is not only lossless, but also at least one of the projections inherits some, possibly all, attributes of R. The result aimed on is that the lossless decomposition possibly does not cost us the logical navigation through the projections, unlike for the original theorems. We leave for the future eventual restatements of many other rules aimed on best schemes, e.g., the Rissanen's theorem.

The major gain that will appear below is that, for the <u>same size optimal schemes</u> for a DB, the one using SIRs effectively spares the discussed logical navigation. More precisely, as we'll show the optimal SIR scheme will be always as follows:

(a) The projections resulting from original Heath's and Fagin theorems applied to stored relations become bases of SIRs resulting from the restated theorems or remain the same.

(b) In the absence of MVDs, no restated decomposition creates the logical navigation through the projections.

(c) Otherwise, a restated decomposition removing an MVD still spares or at least reduces the logical navigation for some queries addressing the projections, but not to all.

(d) The latter result should concern most of real-life queries.

Indeed, first, the Heath's theorem states, we recall, that for any stored relation ABC (A, B, C) and an FD A -> B, the decomposition AB (A, B) and AC (A, C) is lossless. That is: ABC (A, B, C) = AB (A, B) Join AC (A, C). In practice, as known well, we may have several choices for A, B and C. As every decompositions doubles A, for stored relation ABC, it is wise to choose A with fewest attributes, at least for this reason. Likewise, A should be the primary key of AB. B does not depend on any proper subset of A then and AB is in 2NF at least. Also, for reasons previously invoked, we should hunt for the largest B. We may end up nevertheless with AB not in 3NF at least hence B may get decomposed in turn, etc. With all the principles stated in mind, we restate the theorem for ABC being a stored relation or an SIR, as the decomposition into AB (A, B) and AB$^I$C (A, B$^I$, C), where B$^I$ is the IE: B$^I$ (select B from AB where A = AB$^I$C.A). This decomposition is also into two schemes and clearly lossless. But, while AC was a stored relation, AB$^I$C is an SIR with base AC. This decomposition is possible only for the SIR-model. Unlike the original one, while it conserves AC as a projection of ABC, it also preserves the original attributes A, B, C together. It avoids thus, as promised, the logical navigation to queries selecting B and C.

Next, the Fagin's theorem also states that in presence of MVD A ->> B | C in the presumably stored relation ABC (A, B, C), its decomposition into AB (A, B) and AC (A, C) is lossless. Now, suppose B' being a (perhaps empty) subset of B such that A -> B' and let C' be a (perhaps empty) subset of C, where A -> C'. Actually, we may about always expect either B' or C' non-empty, but not both, as in the example that follows. We restate the theorem as follows. Suppose ABC a stored relation or an SIR. The decomposition creates ABC$^I$ (A, B, C$^I$) and ACB$^I$ (A, B$^I$, C) where (i) C$^I$ is the IE C$^I$ (select C' from AC where AB.A = A) defining C' thus and (ii) B$^I$ defines B' as B$^I$ (select B' from AB where A = AC.A). C$^I$ and B$^I$ avoid the logical navigation for any query to B and C' or to B' and C in the projections. Only a query to B/B' and C/C' still needs it. We thus do not avoid completely the logical navigation that the decomposition creates. But we do limit it to fewer queries. Furthermore, as it will appear the remaining queries should usually have the logical navigation through the final optimal scheme of the DB partly limited. Notice that the restated theorem again conserves each original stored projection as is or as the base of one of the SIR projections.

On these bases, the generic schema generation algorithm for SIRs is quite analogous to that for the stored relations only. More precisely, U remains the starting point, except that it may have IAs upfront. From there, we perform the same, wisely chosen, successive decompositions eliminating MVDs and "annoying", i.e., anomaly creating, FDs. However, at each step, we now use a restated theorem instead. If we face both dependencies, the restated Fagin's theorem works first. We naturally end up with the same stored relations, hence the same size scheme, but also with less logical navigation, as claim (b) states. If there are no MVDs, we remove the discussed logical navigation entirely, as claim (a) states. Finally, the rationale for claim (c) is that in a real-life DB, MVDs are rare with respect to annoying FDs. Also, B' or C' usually have several attributes, unlike B/B' or C/C'. Even for a decomposed MVD, most queries to the projections should then normally be logical navigation free as well.

The following example illustrates all the debated points.

Example 8. The biblical S-P1 scheme results from Heath's theorem only. Similar schemes are countless in practice, as widely known. Our scheme in Example 1 would need the restated Heath's theorem only. To illustrates also the restated Fagin's one, we modernize S-P1. Each supplier has now one or more contact email addresses. Each address may serve for any inquiry about the supplies or the supplier itself. Each address is the value of new stored attribute EMAIL. Every address is for one and only one supplier. We redesign the S-P scheme under SIR-model accordingly. We call the result S-P4.

We start optimistically with the universal relation U, [MUV84]. In short notation we have:

U (<u>EMAIL</u>, S#, SNAME, STATUS, SCITY, <u>P#</u>, PNAME, COLOR, WEIGHT, PCITY, QTY).

Notice the necessarily different names for the supplier and part cities, unlike in S and P of S-P1 or S-P2. U is potentially the optimal stored relation for S-P4, unless proven otherwise. What's easy, since EMAIL already introduces the MVD: S# ->> EMAIL | (SNAME, CITY, STATUS, P#...QTY). U is not in 4NF thus. Regretfully, U cannot be the optimal S-P1 scheme. We have to decompose it. We have MVDs and obviously FDs. We start with the restated Fagin's theorem. The decomposition creates two relations:

SE (S#, <u>EMAIL</u>, I_SP (select SNAME, STATUS, SCITY As CITY from SP Where SE.S# = S#)), SP (<u>S#</u>, SNAME, SCITY, STATUS, <u>P#</u>...QTY).

SE is now an SIR where I_SP is the IE denoted above as $C^I$. Its base (S#, EMAIL) would be the stored relation for the original Fagin's decomposition. SP is the same for both decomposition. We now have thus C' = (SNAME, STATUS, CITY) and B' = $\varnothing$. SE is in the (restated) BCNF. It would not be if any of its IAs, e.g., SNAME, was a stored attribute. The IAs of SE spare the logical navigation to any queries to EMAIL and to any of IAs in I_SP. Otherwise, these queries would navigate over SE and SP. In contrast queries selecting emails and an attribute in SP that was not inherited in SE would still need to navigate, i.e. would require the SE join SP clause. We come back to these queries later on, showing that practical ones should require lesser navigation anyway, backing up our earlier claim.

SE has no more MVDs, hence it is also in 4NF. SP has no more MVDs neither. But, is not in (restated) BCNF (hence neither in 4NF). The restated Heath's theorem applies. For all the already discussed reasons, we choose the following decomposition, with S# as A, in particular since it is a single attribute key:

S (<u>S#</u>, SNAME, STATUS, CITY), SP (<u>S#</u>, <u>P#</u>, PNAME...CITY, QTY, I_S (Select*/S# From S Where S# = SP.S#)).

In the projections, we could conveniently rename PCITY and SCITY to simply CITY. The projection SP is again an SIR, with I_S being $B^I$. The related supposedly stored attributes of the decomposed SP, remain thus preserved in the projection SP, in the form of being as inherited from S. Notice that this does not change anything for SE scheme. S remains the stored relation, as for the original decomposition. It is in BCNF. SP however still isn't in restated BCNF. Its projection on the stored attributes indeed isn't in BCNF in SRV-model, given the FD : P# -> PNAME, COLOR, WEIGHT, PCITY. We thus apply the restated Heath's theorem again. One gets SP decomposed to:

P (P#, PNAME, COLOR, WEIGHT, CITY) and SP (P#, S#, QTY, I_S (Select*/S# From S Where S# = SP.S#), I_P (Select */P# From P Where P# = SP.P#)).

Now S-P4 has every relation in BCNF, hence in 4NF, as there are no more MVDs. The optimal scheme is as follows. We underlined the primary key stored attributes.

S (<u>S#</u>, SNAME, STATUS, CITY),

P (<u>P#</u>, PNAME, COLOR, WEIGHT, CITY),

SE (S#, <u>EMAIL</u>, I_SP (select SNAME, STATUS, SCITY As CITY From SP Where SE.S# = S#)),

SP (<u>P#</u>, <u>S#</u>, QTY, I_S (Select*/S.S# *From S, SP Where S.S# = SP.S#), I_P (Select*/P.P# From P, SP Where P.P# = SP.P#)).

Notice that SP scheme is that of S-P2 from Example 1, except for the use of '*/' clause instead of the original lists. That is why the S-P2 scheme is the optimal one as well. Because of IEs, most practical queries is now clearly logical navigation free. However the already signaled queries to SE and SP are not. Some of these queries, e.g., select every P# supplied by supplier with given EMAIL, seem without practical interest. Clients in practice need also names. Then, the restated decomposition still reduces the logical navigation by two joins otherwise necessary, i.e., SE with S to get SNAME and SP with P to get PNAME. We may thus reasonably expect at least some logical navigation spared for practical queries to SE and SP together and for most of such queries to projections of a decomposed MVD in general.

Also, if we did not start decomposing U with the Fagin's theorem, but with Heath's one, the result would be the sub-optimal we spoke about. Indeed, the first decomposition of SPE could use the FD : EMAIL –> S#, leading to:

SE (S#, EMAIL), SP' (EMAIL, SNAME…P#... I_SE (Select S# From SE Where EMAIL = SP'.EMAIL))

SE is again in BCNF. But now, SP' is also free from any MVD, hence we do not need Fagin's decomposition for it neither. However, SP' isn't (yet) in restated BCNF. Through successive restated Heath's theorem decompositions, the final scheme for S-P would be:

S' (EMAIL, SNAME, STATUS, CITY, I_SE (Select S# From SE Where EMAIL = SP'.EMAIL)),

SE (S#, EMAIL),  P (P#, PNAME, COLOR, WEIGHT)

SP' (P#, EMAIL, QTY, I_SE (Select S# From SE Where EMAIL = SP'.EMAIL), I_S' (Select*/S.EMAIL, From S' Where EMAIL = SP'.EMAIL), I_P (Select*/P.P# From P Where P# = SP'.P#))

Now, if a supplier had $m$ email addresses on the average, S' and SP' would have each $m$ time more stored values on the average than, respectively, S and SP. We have more stored values than before, i.e., a sub-optimal result, as predicted.

Finally, suppose for S that we calculate STATUS as in Example 1. The only change to S would be the IE defining STATUS:

S (S#, SNAME, STATUS (Select INT(SUM(QTY)/100) FROM SP_B WHERE S.S# = S#), CITY). @

## 4.   Implementing an SIR DB

### 4.1   Basic Processing Scheme

As said already, the most practical way towards the SIR-model enabled DBS, is to transparently manage an SIR DB by an existing (kernel) SQL DBS. One way is to create the *SIR-layer* managing the SIR DB through calls to the kernel services, Figure 5. For the kernel, SIR-layer appears as any client (user or application). SIR-layer processes every DDL or DML statement for an SIR DB through the internal generation of these for the kernel. It's obviously useful to have the SQL syntax at the SIR-layer as compatible as possible with the kernel SQL dialect. Below, we presume the total immersion of the kernel syntax in the enhanced one.

In particular, for the Create Table statement received, SIR-layer should determine the type of the relation to create. Without an IE, it is a stored relation. SIR-layer should push down the statement as is. Generally, the SIR-layer should forward as is every statement addressing only stored relations or views.

In turn, the processing of statements addressing SIRs should be clearly more involved. First SIRs obviously need dedicated kernel meta-tables for the IEs. The schemes of these are easy enough to skip details. Let us deal first with the creation of an SIR, say R (B, V) again.  The SIR-layer recognizes this one by the presence of IEs in the Create Table R statement it gets from the client. The simplest design seems to represent R in the kernel as the full view, also named R there. This one defines the virtual table with the same attributes and values as R. The left to right order of the attributes in the Select clause of the full view is furthermore that of their top-down and left to right appearance in the Create Table R statement at SIR-layer, e.g., as at Figure 3 and Figure 4. In other words, it is a view created as if the kernel could execute Create View R_V As Select * From R; renaming then R_V to R. SIR-layer simply forwards afterwards every incoming query to the kernel basically as is. The kernel takes care, in particular, of the query evaluation. SIR-layer avoids this complex burden.

However, the kernel cannot create a full view of an SIR directly as above. SIRs are indeed unknown within. Below, we propose that SIR-layer implements the full view R through (i) stored relation B and (ii) a sequence of view(s) creating progressively V and adding IAs within to those selected from B. Each view results from the previous ones and from one IE. The view R is the last one. We now define this processing more formally. For sake of comprehension, we remain nevertheless still rather informal. We qualify the overall result of *basic (processing) scheme* for SIRs, BPS in short.

As we have seen, an IE may define IAs through (a) Select expression or (b) through value expressions only, e.g., for WEIGHT_KG, WEIGHT_T or RANK. SIR-layer starts with the rewriting of some IEs into a *canonical* form. The rewriting depends on IE. Let $I'$ be an IE and $I$ the same one but rewritten into the canonical form. Let us start with an $I'$ of kind (a), hence written as I' (Select A From F' Where W'). Here A denotes all the elements of the Select clause. F' and W' are the contents of From and of Where clauses. Every element in A is an attribute name or a named value expression perhaps with an aggregate or scalar function, or an attribute name with an alias. In general, it could be also a named subquery. We do not see the need for this case in an IE at present. $W'$ contains the recursive (inner) join clause(s), e.g., X.J1 = R.J2 for a single one. Then, with the exception of $I'$ defining a single IA through a value expression with an aggregate function, e.g., STATUS, SIR-layer rewrites W' into W without this join and, for each recursive join clause, rewrites F' into F = R Left Join F' on X.J1 = R.J2. E.g., the IE I_S, Figure 3, would become $I$:

$I$ = I_S (Select SName… From SP Left Join S on S.[S#] = SP. [S#]) ;

If there are several recursive joins in W', then the outer join clauses nest in F as usual. Notice that a recursive join is in general a $\theta$- join with perhaps $\theta \neq$ '=', e.g., we had $\theta$ = '>' for RANK.

Each exceptional $I'$, e.g., STATUS, is in the canonical form by default for SIR-layer. Same for every IE of type (b) with several value expressions. In contrast, SIR-layer rewrites every IE $I'$ of this type stated as I' As (*V*) to the canonical form of $I$ = I' (*V* As I'), i.e., we now have A = *V* As I'. For instance: $I'$ = WEIGHT_T As (WEIGHT_KG/1000) becomes WEIGHT_T (WEIGHT_KG/1000) As WEIGHT_T). Next, consider for some standard IE $I$ in relation R with R' defined as in Section 2.1, the following pseudo SQL expression:

(V1) Create View *R* As Select R'.*, A From F Where W;

We recall that we do not use the name R' operationally, only as pseudo SQL notation to designate all the attributes outside the sub-tuple created by $I$. If $I$ is unique in R, *R'* is B. For any $I$, *R* is a view with the same attributes and tuple values as the full view. However, the attributes may be in the order different from that of Select * from R. *R* is thus only relationally equivalent to the full view. The original order may matter to an SQL user. If $A_1…A_n$ denote all and only attributes of *R'* and of $I$ in their original order in Create Table R, the full view R may result from one more statement:

(V2) Create View R As Select $A_1…A_n$ From *R*;.

However, (V2) creates a mapping from R to *R*, to deal with during query evaluation. It should be more efficient operationally, to rather generate (V2) directly as (V1), but with the attributes $A_1…A_n$ in order, i.e., as

(V3) Create View R As Select $A_1…A_n$ From F Where W.

Formula (V3) may suffice operationally if $I$ is unique in R. The general case is however that of multiple IEs. Every R' is then a view, depending not only on B, but also on all but the IE chosen for (V3). The formula cannot help operationally. BPS creates therefore R more generally through the following sequence of partial views, ending up with the full one.

Let $I_1...I_n$ be the IEs in R scheme, standardized and numbered in the order of their evaluation. The order should let each IE to find all the attributes it refers to in the view resulting from the previous IEs. For instance, for our example of WEIGHT_KG and of WEIGHT_T, created by two IEs, the view with WEIGHT_KG should be created first. Let it be $R_0$ = B, operationally named R_B in the kernel. Next, let $R_1...R_n$ be the views produced each respectively by the evaluations of $I_1, I_2...I_n$, as follows.

- If $I_i$ has Select expression and $I_i$ is not the exception we spoke about, then SIR-layer creates the view as:

(V4) Create View $R_i$ As Select $R_{i-1}$.*, $A_i$ From $F_i$ Where $W_i$ ;     /* $i$ = 1…n

Each left join in each (canonical) $W_i$ is adjusted here so to refer to $R_{i-1}$ as R, when the prefixing by relation name is mandatory. If $I_i$ is an exception, e.g., STATUS, then its content becomes a subquery named $I_i$. Namely, SIR-layer generates:

(V5) Create View $R_i$ As Select $R_{i-1}$.*, (Select A From F Where W) As $I_i$ From $R_{i-1}$ ;

If $A_i$ in $I_i$ is one or more (canonically) named value expressions only, then SIR generates:

(V6)  Create View $R_i$ As Select $R_{i-1}$.*, $A_i$ From $R_{i-1}$ ;        /* $i$ = 1…n

The full view of R results from eventual reordering of $R_n$ through (V3).

Indeed, view $R_1$ is the join of B and of $I_1$ through the recursive join clause or is B.* with additional virtual attributes. Likewise $R_2$ joins all the tuples of $R_1$ that serves as R' in turn, with the matching values of IAs of $I_2$ or a null $I_2$ tuple. Or it adds new virtual attributes. Etc.

Practically, to process a Create Table R statement, the SIR-layer starts with the standardization of all the IEs needing it into the canonical form. Then, it requests from the kernel to create the stored relation R_B. Next, it creates the view $R_1$ named, say, R_1, then $R_2$ as R_2 etc. until the view $R_n$. SIR-layer verifies whether this one would be a full view. If so, it renames it R. Otherwise it applies (V2) and replaces $R_n$ with view R.

Example 9. (1) We submit to SIR-layer S-P2 scheme at Figure 3 to create. SIR-layer applies BPS and finds no IEs in Create Table S and Create Table P. It passes each statement as is to the kernel. It finds IEs in Create Table SP. It proceeds according to rules (i) to (V6) above. It does not find in I_S any attribute created by I_P and vice versa.  It therefore chooses $I_1$ = I_S and $I_2$ = I_P. Consequently, according to BPS, the SIR-layer issues the statements:

Create Table S…   /* Usual creation of a stored table with the attributes at Figure 3.

Create Table P…  /* Idem

Create Table SP_B…  /* From all and only stored attributes of SP at Figure 3.

Create View SP_1 As select SP_B.*, SNAME, STATUS, CITY As SCITY From SP_B Left Join S On SP_B.S# = S.S# ;  /* I_S is rewritten to its canonical form

Create View SP As select SP_1.*, PNAME, COLOR, WEIGHT, CITY As PCITY From SP_1 Left Join P On SP_1.P# = P.P# ;   /* I_P is in the canonical form, SP_2 is SP directly.

(2) Suppose that the user wishes to create the DB S-P3, Example 2. For relation S that became an SIR, we recall, BPS generates the following statements for the kernel:

Create Table S_B…  /* With S#, SNAME, CITY

Create View S_1 As Select S_B.*, (Select Int ( SUM (QTY) / 100) From SP_B Where S_B.S# = S#) AS STATUS  FROM S_B;    /* The STATUS IE, see Example 1.3, became a subquery in Select clause.

Create View S As SELECT S_1.S#, S_1.SName, S_1.STATUS, S_1.City FROM S_1;

Then, for relation P, BPS generates $I_1$ = WEIGHT_KG and $I_2$ = WEIGHT_T, since the latter refers to the former. Consequently, BPS issues the following four statements for P to the kernel:

Create Table P_B…

Create View P_1 As Select P_B.*, WEIGHT/2.1 As WEIGHT_KG From P_B;  /* $I_1$ is the canonical WEIGHT_KG.

Create View P_2 As Select P_1.*, WEIGHT_KG/1000 As WEIGHT_T From P_1;  /* $I_2$ is the canonical WEIGHT_T.

Create View P As Select P#, PNAME, COLOR, WEIGHT, WEIGHT_T, WEIGHT_KG, CITY From P_1; /* Full view of P after attribute reordering in P_2.

Finally, for SP, BPS generates same statements for the kernel as for S-P2, except that I_P, hence also SP, include now also SPWEIGHT_T and SPWEIGHT_KG, we recall.@

Figure 5 illustrates Example 9.2. The figure recalls that S-P3 has only SIRs. The SIR-layer shows these as rectangles. Each size reflects the number of tuples and the number of attribute values per tuple as seen by the client. This perception corresponds to Figure 4, augmented with the IAs proper to S-P3. The lower part shows under the same convention the stored relations and views possibly implementing S-P3 over some current DBS. We represent the views created by BPS in the bottom up order. A rectangle name reflects a view name and the IE used for its creating.  The full view P restoring the original order is shown as P(*).

View representations at the figure illustrate the IEs creating them. Each rectangle length is the same as for the SIR-layer rectangle. But it is not so for the width, for the rectangles representing the stored relations especially.

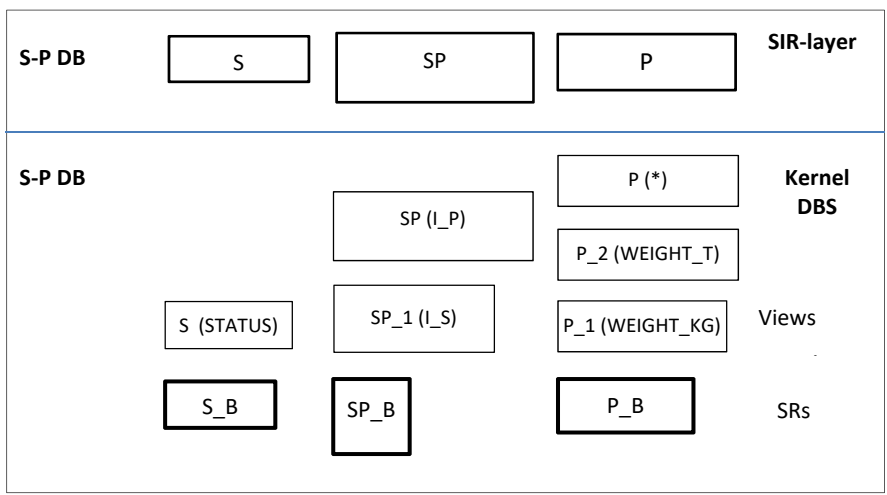

Figure 5 S-P3 DB. Above: SIRs. Below: Views and stored relations (SRs) possibly implementing S-P3 within an existing kernel DBS.

As the figure shows, for the three SIR schemes, the BPS would generate nine relational schemes in DBS. Three would be the stored relation schemes. Hence, they would define the relational conceptual scheme of S-P3 in the SRV-model. The views would help the queries to S-P3 to be less procedural, avoiding some logical navigation, value expressions and perhaps additional mandatory auxiliary views, as we discussed. Without SIRs, the user or DBA wishing simpler queries would need to basically create all these six views manually. The virtual attributes could spare both views for P and for P only. The currently available implicit join capabilities would spare nothing. As said, the SIR-layer could alternatively generate only a more complex but single view per SIR. It would lead to the minimum of two schemes per SIR in an actual DBS, or even only one if SIR-layer generates the virtual attributes. But, the single-view strategy could end up a bad idea for multiple IEs, as pointed out. Finally, to appreciate the non-procedurality gain our sample SIRs may bring, perhaps formulate to S-P1, assumed in SRV-model, the query equivalent to the following simple one to S-P3: Select SNAME, SUPPLIERS, PNAME From SP Where STATUS > 2 And SPWEIGHT_T > 1 ORDER BY STATUS@

With respect to the other DDL statements for SIRs, the Alter Table and Drop Table also require more processing than their kernel counterparts. For Alter, the kernel may in particular create/drop subqueries when an SIR gets/loses IAs. E.g., see how partial views for S get altered if one adds RANK to S.

SIR-layer should be implemented in some host language, obviously calling the Embedded SQL interface of the kernel. This is a future work. In the meantime, [L6a] backs up Example 9 with a manual simulation on MsAccess as the kernel. For each simulated SIR, a stored MsAccess table is its base. The MsAccess stored queries simulate all the views the BPS would create. The result lets an MsAcces user to practically appreciate advantages of SIRs, through queries to the full views. One may also modify the views, e.g., to experiment with everything above discussed. As an easy bonus, one may experiment the QBE interface, generate forms, graphics, etc. In sum, one may play with all nice capabilities of MsAccess that made it so popular, almost as if they were designed for SIRs as well.

Performance-wise, the kernel storage for an SIR is in practice the one for its base. Full view storage is negligible provided the view is not materialized, as we suppose. As shown, the optimal scheme with SIRs has the same size and the same stored attributes as the optimal one for SRV-model. Hence, the storage for the values of these SIRs within the kernel is the same. The optimal DB with SIRs for some application should cost thus negligibly more than the optimal DB with the stored relations only for the same goal. Next, since the SIR-layer passes each query as is to the kernel, its query evaluation overhead is negligible as well. Whether any "competitor" at SIR-layer provides queries with a more advantageous evaluation within the kernel, is currently one more anyone's guess. Altogether, perhaps surprisingly, the enticing capabilities of SIRs appear at present practically without operational overhead.

## 4.2 Enhancing BPS

Examples show that BPS should usually suffice. As usual for a DBS component, one can nevertheless enhance it. Below we discuss a few examples. Some enhancements aim on kernels with specific capabilities only. Whether any enhancements will actually prove practical enough to get implemented, is anyone's guess at present. We leave the exhaustive investigation of BPS enhancements for the future work.

First goal for BPS enhancements is fewer views $R_i$. The rationale is the speed-up of queries through fewer view mappings. One easy rule is to avoid generating the full view $R_n$ if view $R_{n-1}$ is already an equivalent one. Instead, Enhanced BPS can then generate the full view directly from $R_{n-2}$. It suffices to take into the account the attribute order as well. In Example 9.2, SIR-layer would then generate view P directly from P_1, gaining one mapping.

Another easy rule aiming at the same purpose is to collapse the IEs defining value expressions only into a single one. In Example 9.2 this could spare P_2. With both rules, view SP would even replace P_1. The kernel would end up with two view mappings less than for BPS. The kernel supporting virtual attributes could even replace SIR P with a stored (only) P. Notice, however that in turn, for Example 9.1, the discussed rules would surely complicate BPS, but would gain strictly nothing. The complication would even naturally little slow down then the SIR-layer.

Another issue is that of the circular references. Examples show enhancements to BPS avoiding these in the DBS, despite having some in the SIR DB scheme. They point to rewriting rules of the affected SIRs or views during the actual scheme generation. The goal here is the DBA comfort, not the usual query execution speed-up. E.g., one could let in this way the DBA of S-P3, to declare STATUS referring simply to SP as in S-P2, Example 1.3, despite the induced circular reference between S and SP. The enhanced BPS could automatically rewrite it into the IE referring to SP_B, as discussed in Example 1.4.

## 5. Conclusion

Stored and inherited relation, (SIR), appears a useful construct for a relational DB. An SIR may be free of anomalies of a stored (only) relation with the same attributes. Through the inherited attributes, an SIR scheme may be also more accurate as the conceptual model. SIRs alleviate in this way the well-known limitations of stored relations, the dark side of the normalization. The popular ER model, proposed precisely because of these limitations, becomes rather useless.

The main practical gain is less procedural queries.  The SIR-layer appears from this stance as a higher level interface to the relational DBs. At first, through the reduced logical navigation for the optimal conceptual scheme with the same stored relations. Next, SIRs may avoid managing numerous additional view schemes, often necessary at present for user's comfort, especially to hide complex value expressions. The inherited attributes of an SIR generalize also for both goals the already popular view saver practices of virtual attributes and of implicit joins. Finally, to implement the SIR-layer using an existing DBS looks rather easy and without operational overhead in practice.

The future work should start thus with the implementation of SIRs over a popular kernel DBS along the gross architecture we defined. Depending on the kernel's actual capabilities, it may be wise to include some of enhancements to BPS we have discussed. But even without these, the result should be the win-win deal.  Better late than never, the existing DBSs should get improved accordingly.

On the theoretical side, the design rules for SIRs based on restated NFs and Heath's and Fagin's theorems appear about as easy as the current ones. However, the decompositions based on these two theorems exclusively, are only the tip of the iceberg of known proposals. Future work could adapt those proposals to SIRs as well, especially the proposals for the lossless decomposition using outer joins, [JS90].

One may also observe further that all the three constructs, i.e., SRs, IRs and SIRs root in a common 1NF construct. One could call it *relation with stored <u>or</u> inherited attributes*, or *stored or inherited relation*, say (SoIR) in short. As the name hints, a SoIR may be a stored one, or a view or an SIR. The construct may look esoteric. Observe however that our Create Table for an SIR is in fact that one of a SoIR, except for a view. Our design rules apply to SIRs, but to SoIRs as well. Summing up, whether one sees our work as on SIRs or on specific SoIRs is just a matter of taste.

SoIR construct roots itself in a still more general 1NF construct that one may call *relation with stored or inherited (attribute) values*. The idea, outlined already in [LKR92], is of attributes possibly mixing stored and inherited values. A stored value of such an attribute whenever present, overrides the eventually inherited value. This may be obviously practical.  For instance if color of a part in P is green, while an S-P2 user getting a supply involving the part in SP rather sees it as light green, for any reasons, then the update overriding the color in the related SP tuple could bring the global happiness. Future work could explore that issue as well.

Finally, most of major DBSs are now interoperable multidatabase systems, [LA86]. SIRs with multidatabase IEs seem attractive as well.